# Microscopic mechanisms of flexoelectricity in oxide membranes

*Harikrishnan KP[1], Varun Harbola[2,3,4], Jaehong Choi[1], Kevin J. Crust[3,4], Yu-Tsun Shao[1,5], Chia-Hao Lee[1], Dasol Yoon[6], Yonghun Lee[3,7], Gregory D. Fuchs[1], Cyrus E. Dreyer[8,9], Harold Y. Hwang[3,7], David A. Muller[1,10],\**

[1.] *School of Applied and Engineering Physics, Cornell University, Ithaca, New York 14853, USA*
[2.] *Max Planck Institute for Solid State Research, 70569 Stuttgart, Germany*
[3.] *Stanford Institute for Materials and Energy Sciences, SLAC National Accelerator Laboratory, Menlo Park, CA 94025, USA*
[4.] *Department of Physics, Stanford University, Stanford, CA 94305, USA*
[5.] *Mork Family Department of Chemical Engineering and Materials Science, University of Southern California, Los Angeles, CA, 90089 USA*
[6.] *Department of Materials Science and Engineering, Cornell University, Ithaca, New York 14853, USA*
[7.] *Department of Applied Physics, Stanford University, Stanford, CA 94305, USA*
[8.] *Department of Physics and Astronomy, Stony Brook University, Stony Brook, New York 11794, USA*
[9.] *Center for Computational Quantum Physics, Flatiron Institute, 162 Fifth Avenue, New York, New York 10010, USA*
[10.] *Kavli Institute at Cornell for Nanoscale Science, Ithaca, New York 14853, USA*
*\* Corresponding author : david.a.muller@cornell.edu*

## Abstract

Modern electromechanical actuators and sensors rely on the piezoelectric effect that linearly couples strain and electric polarization. However, this effect is restricted to materials that lack inversion symmetry. In contrast, the flexoelectric effect couples strain gradients to electric polarization, and is a universal property in insulating materials of arbitrary symmetry. Flexoelectricity becomes prominent at the nanoscale from the inverse scaling of strain gradients with material dimensions. Here, we measure the strain-gradient-induced structural distortions in strontium titanate using multislice electron ptychography. This technique enables reliable picometer-scale measurements of the dominant oxygen-titanium distortions, correcting for artifacts that limited conventional imaging methods. This enables us to directly measure the sign of the net ionic contribution to the flexoelectric polarization. Guided by the experimental measurements, first-principles calculations show how the sign and magnitude of the bulk contribution to the flexoelectric coefficient in strontium titanate can be switched by tuning the strain state. Hybridization between the optical soft phonon and acoustic phonon modes drives this transition, yielding a large response and a polarity switch across the resonance. This strain-dependence might explain the sign discrepancy and orders of magnitude variation in the values of previously reported flexoelectric coefficients for strontium titanate. As the strain state of curved membranes can be tuned, our approach also suggests an approach to engineer nanoscale flexoelectric polarization using strain as a control parameter.

## **Main**

Piezoelectricity, that describes the linear coupling between mechanical strain and electric polarization, is the most mature technology for electromechanical transduction, enabling applications from quartz crystal oscillators to precision actuators. However, piezoelectricity is limited to non-centrosymmetric materials and does not scale well as device dimensions shrink, posing challenges for nanoscale devices and on-chip integration. At the nanoscale, a promising alternative is the flexoelectric effect[1–5] that couples strain gradients and polarization and is described by a 4$^{th}$ rank tensor:

$$P_i = \mu_{ijkl} \, \partial_j \varepsilon_{kl} \tag{1}$$

Here, $P_i$ represents the polarization, $\mu_{ijkl}$ is the flexoelectric tensor and $\partial_j \varepsilon_{kl}$ is the strain gradient. Although flexoelectricity is generally a miniscule effect, it becomes more prominent at the nanoscale due to the inverse scaling of strain gradients with material dimensions. Unlike piezoelectricity which is symmetry-restricted to the 20 non-centrosymmetric point groups, flexoelectricity is universal in materials of all symmetry as strain gradients themselves break centrosymmetry.

Despite being a universal property, flexoelectricity remained largely underexplored for many years, partly due to phenomenological estimates[1] predicting extremely small flexoelectric coefficients (1–10 nC/m). These values implied that strain gradients exceeding $10^6$ m$^{-1}$ would be necessary for any significant impact on the electrical properties of a material. However, beam bending experiments by Ma and Cross[6–8] in the early 2000s revealed flexoelectric coefficients that were 3-4 orders of magnitude larger than these initial estimates, reigniting interest in the field. Nevertheless, achieving such high strain gradients in bulk still remains challenging due to the brittle nature of most dielectric materials. Although substantial strain gradients can be stabilized by lattice mismatch and strain relaxation in thin films[9,10] or at defect sites like crack tips[11] or dislocations[12,13], the degree of tunability is severely restricted. The ability of free-standing membranes[14,15] to accommodate substantial strains[16–19] due to their enhanced elastic compliance, combined with their excellent flexibility for manipulation into different shapes and geometries[20,21], has unlocked new opportunities to explore flexoelectric effects. The non-uniform topologies of free-standing membranes can stabilize large strain gradients, and the resulting polarization from the flexoelectric effect can be comparable to, or even larger than that in ferroelectric thin films[22]. Here, we perform a fundamental investigation into the microscopic nature of strain gradients and flexoelectric polarization in bent membranes of strontium titanate.

Conventional scanning transmission electron microscopy (STEM) techniques are either completely insensitive to light atoms or prone to systematic errors in their measured positions, due to multiple-scattering effects and mistilt-induced artifacts[23–26]. This problem can now be addressed through multislice electron ptychography (MEP)[27,28] which enables reliable measurement of atomic positions including those of the light oxygen atoms[29–33]. Through direct measurements of atomic displacements from MEP images, we show that the polar distortions are aligned with the strain gradient direction in the bent $SrTiO_3$ membrane, revealing structural distortions dominated by oxygen displacements relative to the Ti sublattice and minimal cation-cation displacements. This observation from MEP is consistent with our first-principles calculations where the oxygen displacements are an order of magnitude larger than the displacements of the Ti atoms when the Sr sublattice is taken as the reference. Moreover, our calculations show that the magnitude and sign of the bulk contribution to the flexoelectric coefficient in $SrTiO_3$ is sensitive to the strain state of the membrane and can potentially be switched by tuning either the in-plane or out-of-plane strain. Thus, the flexoelectric polarization can be switched not only by flipping the direction of the strain gradient, but also by controlling the strain state of the membrane. This coupling between the flexoelectric response in $SrTiO_3$ and its strain state is driven by the hybridization between the optical soft phonon mode in $SrTiO_3$ and the acoustic phonon modes[34], creating a resonance with significantly enhanced response and a switch in the sign of the response across the resonance. We also note that the strain-induced ferroelectric phase transition in $SrTiO_3$ due to the condensation of this polar soft mode has been extensively studied[35–38]. This ferroelectric polarization can be either in-plane for biaxial tensile strain or out-of-plane for biaxial compressive strain, with the critical strain magnitude for the transition exhibiting strong temperature dependence. Our findings are crucial for addressing the sign discrepancy and order of magnitude variations in the values of previously reported flexoelectric coefficients as different fabrication approaches can stabilize largely different strains, that in turn can significantly influence the flexoelectric response. The knowledge of structural distortions from MEP is supplemented by analysis of the local electronic symmetry through electron energy loss spectroscopy (EELS). We observe a monotonic trend in the crystal field splitting energy of the Ti $3d$ orbitals along the direction of the strain gradient and discuss this trend in the context of variations in Ti-O bond lengths in response to the strain gradient and flexoelectric polarization.

**Strain-gradient engineering**

We create wrinkles in 6-15 nm thick single-crystal $SrTiO_3$ membranes by draping them over an array of nanofabricated pillars as shown in the SEM image in Fig. 1(a). The membrane forms wrinkles to conform with the 3-dimensional geometry created by the nanopillars as shown in the top part of the SEM image. The bare array of nanopillars can be seen on the bottom left, and an oblique view of a single nanopillar of height 150 nm is shown in the inset. To investigate the atomic structure of these wrinkles, we prepared cross-sectional TEM samples across the wrinkles with a typical site labeled with a yellow box in Fig. 1(a). The high-angle annular dark field (HAADF) STEM image of a wrinkle is shown in Fig. 1(b), revealing a convex bend near the apex and a more gradual concave bend farther away from the apex. Tracking the atomic planes near/away from the apex (Supplementary Fig. 1) shows that their curvature is maximized near the apex.

Figure 1(c) shows a magnified atomic resolution image taken near the apex of the wrinkle at the region marked with a red box in Fig. 1(b). The convex geometry of the bend induces non-uniform stretching and compression of the atomic planes, resulting in tensile strain at the top and compressive strain at the bottom of the wrinkle. This inhomogeneous strain profile across the membrane is evident in the strain map shown in Fig. 1(d), with the measured $\varepsilon_{xx}$ strain component reflecting the changes in the local in-plane lattice constant.

Further, by tracking the strain distribution across different unit cells in each atomic plane, we produce a plot of the strain profile across the membrane in Fig. 1(e). Each datapoint is calculated as the mean of the strain across all unit cells in the same atomic plane, while the standard error of the mean is used as the error bar. The slope of the line gives the transverse strain gradient, calculated to be $\partial_z \varepsilon_{xx} = 2.1 \times 10^6$ m$^{-1}$, where the x- and z- dimensions represent the local in-plane and out-of-plane lattice directions respectively. This value of the strain gradient is 7 orders of magnitude larger than can be stabilized in bulk SrTiO$_3$[39], and results from the greatly enhanced elastic compliance of oxide membranes. The strain gradient magnitude can be increased further by using thinner membranes or by generating wrinkles with pre-strained polymers[20]. Such approaches require careful consideration of the highly non-linear elastic behavior of membranes[17,40,41], or competing strain relaxation mechanisms discussed in Supplementary Fig. 2.

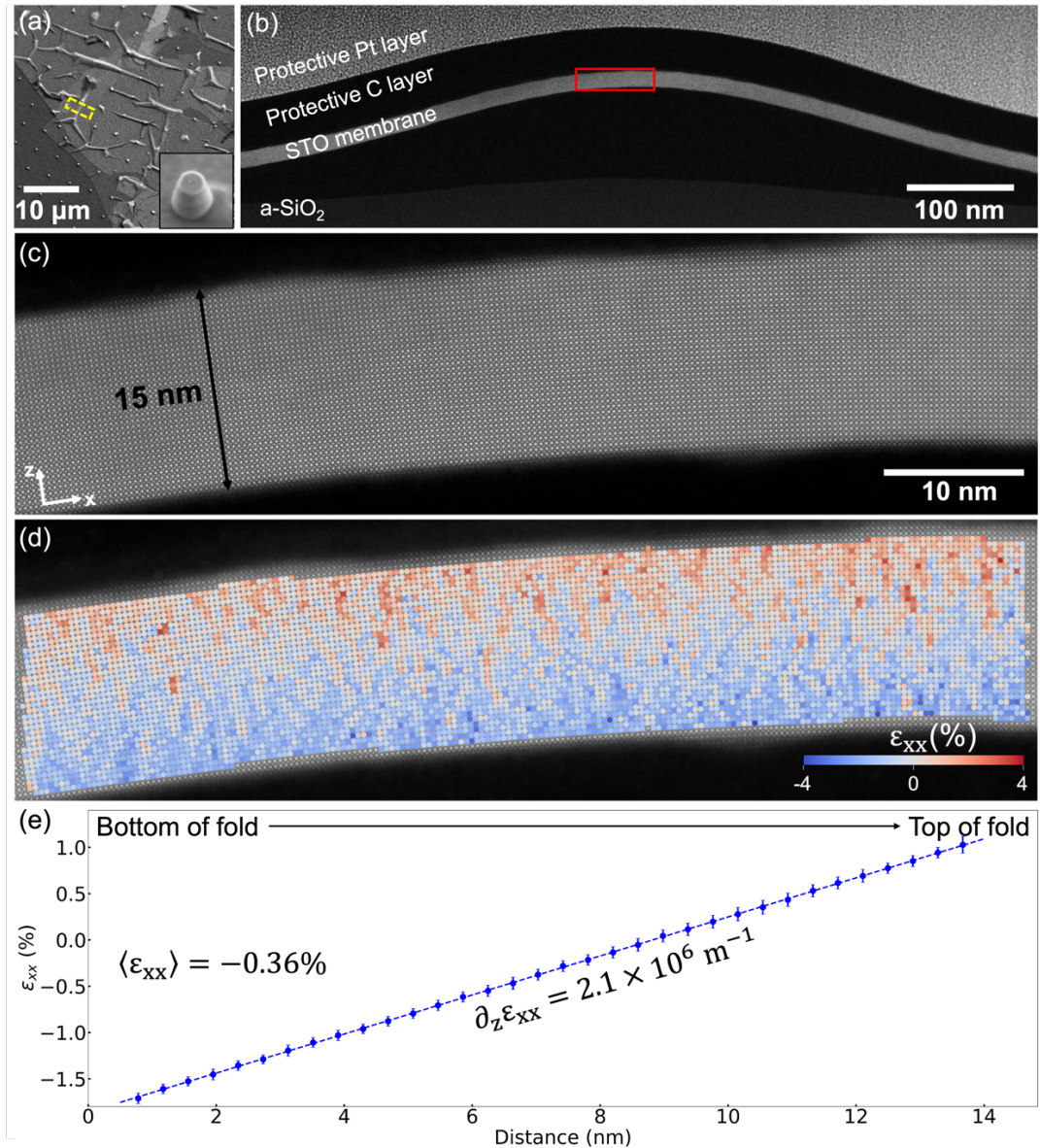

*Fig. 1 | Strain gradient engineering in a free-standing oxide membrane.* **a,** Top view SEM image of a nanofabricated array of $SiO_2$ nanopillars, with a strontium titanate membrane (STO) draped over it forming nanoscale wrinkles. The bare array of nanopillars can be seen on the bottom left, with an oblique view of a single nanopillar shown in the inset on the bottom right. A typical site for preparing cross-sectional sample is labelled with a yellow box. **b,** Cross-sectional HAADF-STEM image of a wrinkle showing the bent $SrTiO_3$ membrane. **c,** Magnified HAADF image at the wrinkle apex acquired from the region marked with a red box in **(b)** shows the curvature of the atomic planes. **d,** Strain map calculated from the atomic resolution image in **(c)** showing the variations in the local lattice parameter (along the in-plane/horizontal direction) reveals the presence of a strain gradient across the membrane. The local strain averaged over each atomic plane is plotted in **(e)** as a function of distance across the membrane and shows a linear increase in strain from the bottom to the top of the fold with a strain gradient magnitude of $2.1 \times 10^6$ $m^{-1}$. Error bars represent standard error of the mean.

In addition to the averaged behavior described above, Fig. 1(d) also reveals subtle non-uniformities in the way the strain is distributed in the membrane with finger-like structures propagating across the membrane that are labeled in Supplementary Fig. 3. Such formation of strain fingers in response to a strain gradient is a universal phenomenon and a similar behavior is observed in the strain profiles of reinforced concrete beams upon mechanical loading[42,43]. The universal nature of these strain fingers[44] spanning nine orders of magnitude in length from nanometer-scale free-standing perovskite membranes to meter-scale reinforced concrete beams is quite remarkable.

**Flexoelectric polarization**

For a bent free-standing membrane where we expect spatial variations in local sample tilt, the accurate characterization of structural distortions is challenging due to systematic errors (induced by mistilts) in atomic positions in conventional STEM techniques like HAADF[23], annular bright field (ABF)[24] and integrated differential phase contrast (iDPC)[26] imaging. Moreover, in thin TEM lamella of polar materials, surface relaxation at the lamella beam entrance and exit surfaces could create an electrically-dead layer that generates image artifacts. Overcoming these artifacts necessitates the use of a technique that can solve for the channeling effects of the electron beam as it propagates through the sample to enable a reliable measurement of atomic positions inside the material, which we achieve with MEP as demonstrated previously[29,45–47]. We account for the local variations in the sample tilt by using a probe position dependent tilt correction in the Fresnel propagator that describes electron propagation within the sample in the ptychographic model[48] (see details in Supplementary Fig. 4).

Using MEP, we demonstrate the direct measurement of the displacements that contribute to the flexoelectric polarization by visualizing relative cation-anion displacements in a wrinkle formed in a 6 nm thick $SrTiO_3$ membrane. The HAADF and MEP images of this wrinkle apex are shown in Supplementary Fig. 5, along with strain maps in Supplementary Fig. 6, used to determine a strain gradient of $2.1 \times 10^6$ m$^{-1}$. Magnified HAADF and MEP images of a 2x2 unit cell region in this wrinkle are shown in Fig. 2(a, b) respectively to compare the imaging capabilities of the two methods. Unlike HAADF imaging, the MEP reconstruction is sensitive to the light oxygen atoms, which as we will see, is crucial for the measurement of the dominant structural distortions. The enhanced resolution of MEP is also apparent from the width of the individual atomic columns.

With no information about the light oxygen atoms, cation-cation displacements (displacement of the Ti atom from the centroid of the Sr atoms for each unit cell) are often used as a proxy to estimate dipole moments in HAADF images. This map of cation-cation displacements is overlaid on the HAADF image in Fig. 2(c) and shows randomly oriented vectors with no general trend. As we will see later with the MEP image, cation-cation displacements are minimal in this sample. Hence, this vector map generated from the HAADF image is just noise, due to the limited precision of the technique in the presence of sample mistilts that can produce false displacements in projection.

In contrast, our MEP implementation corrects for sample mistilts and enables reliable imaging of the light oxygen atoms, allowing precise measurements of the cation-anion displacements to characterize dipole moments. Figure 2(d) shows a vector map of the Ti atom position with respect to the centroid of the oxygen atoms for each unit cell, overlaid on the MEP image. The vector map for the displacement between the centroid of the Sr atoms and oxygen atoms appears similar and is shown in Supplementary Fig. 7. The dipole moments from these structural distortions are aligned with the direction of the strain gradient, with effective cation-anion displacements measured to be $4.2 \pm 1.3$ pm (Ti-O) and $4.0 \pm 1.2$ pm (Sr-O). If we naively use formal valence charges (+2 for Sr, +4 for Ti and -2 for O), we can set a lower bound (as Born effective charges are typically larger) on the polarization of around $6.6 \pm 2$ µC/cm$^2$. From the measured strain gradient and lower bound on the polarization calculated using the formal charges above, the flexoelectric coefficient of the form $\mu_{iijj}$ for the 6 nm thick SrTiO$_3$ membrane is estimated to be 31.43 nC/m (we discuss more rigorous DFT estimates below).

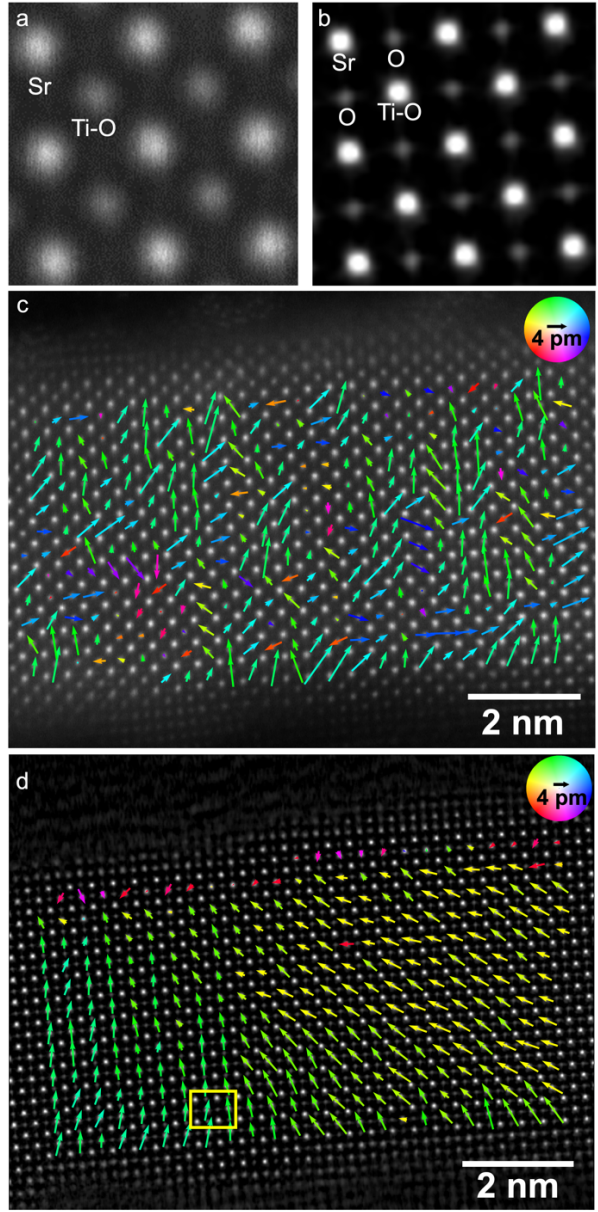

*Fig. 2 | **Atomic scale mapping of the flexoelectric polarization.** Comparison of **a,** HAADF and **b,** MEP image of a 2x2 unit-cell region from a 6 nm thick bent SrTiO$_3$ membrane. In contrast to the HAADF image, the MEP image is sensitive to the light oxygen atoms and offers enhanced contrast and resolution. The different atomic columns are labelled below. **c**, Vector map of the cation-cation displacements (displacement of the Ti atom with respect to the centroid of the Sr atoms for each unit cell) that is often used as a proxy for the polarization, calculated from and overlaid on the HAADF image. This map showing randomly oriented vectors with no general trend is just noise from the limited precision of HAADF imaging in the presence of sample mistilts. **d**, Map of the displacement vectors between the oxygen atom centroid and Ti for each unit cell calculated from and overlaid on the MEP image, showing the dipole moments arising from the flexoelectric effect. The dipoles are aligned roughly along the expected direction of the strain gradient.*

We note that these structural distortions are dominated by displacements of the oxygen atoms with respect to the cation sublattices, with the A-site (Sr) and B-site (Ti) cations remaining centered with respect to each other. These microscopic details are illustrated in Fig. 3(a) which shows a magnified view of 1.5 pseudo-cubic unit cells (extracted from the yellow box in Fig. 2d) with the positions of Sr, Ti-O and O columns measured by 2D Gaussian fitting, marked with red, blue and green solid circles respectively. Dotted lines are drawn through the centers of the cation columns, with respect to which the downward shift of the oxygen positions is clearly visible. A schematic depicting the flexoelectric response of $SrTiO_3$ to strain gradients observed here is shown in Fig. 3(b). The top panel shows 3x3 unit-cells of an undistorted $SrTiO_3$ crystal. The bottom panel shows an exaggerated view of the crystal after a strain gradient is applied to it - with the in-plane lattice constant increasing from the bottom to the top. In response to the strain gradient, the crystal develops polarization in the direction of the strain gradient through the flexoelectric coupling. At an atomic level, the dipole moment develops from a displacement of the oxygen atoms relative to the cations in a direction opposite to that of the strain gradient in every unit cell. As the distortions are dominated by oxygen displacements relative to the cations, tracking cation displacements alone in HAADF images fail to produce an accurate picture of the polarization, instead producing a map of randomly oriented vectors with magnitude close to the noise floor of the measurement precision as shown in Fig. 2(c). We also provide additional proof of the presence of flexoelectric polarization from the breaking of Friedel symmetry in convergent beam electron diffraction patterns collected near the wrinkle apex as illustrated in Supplementary Fig. 8.

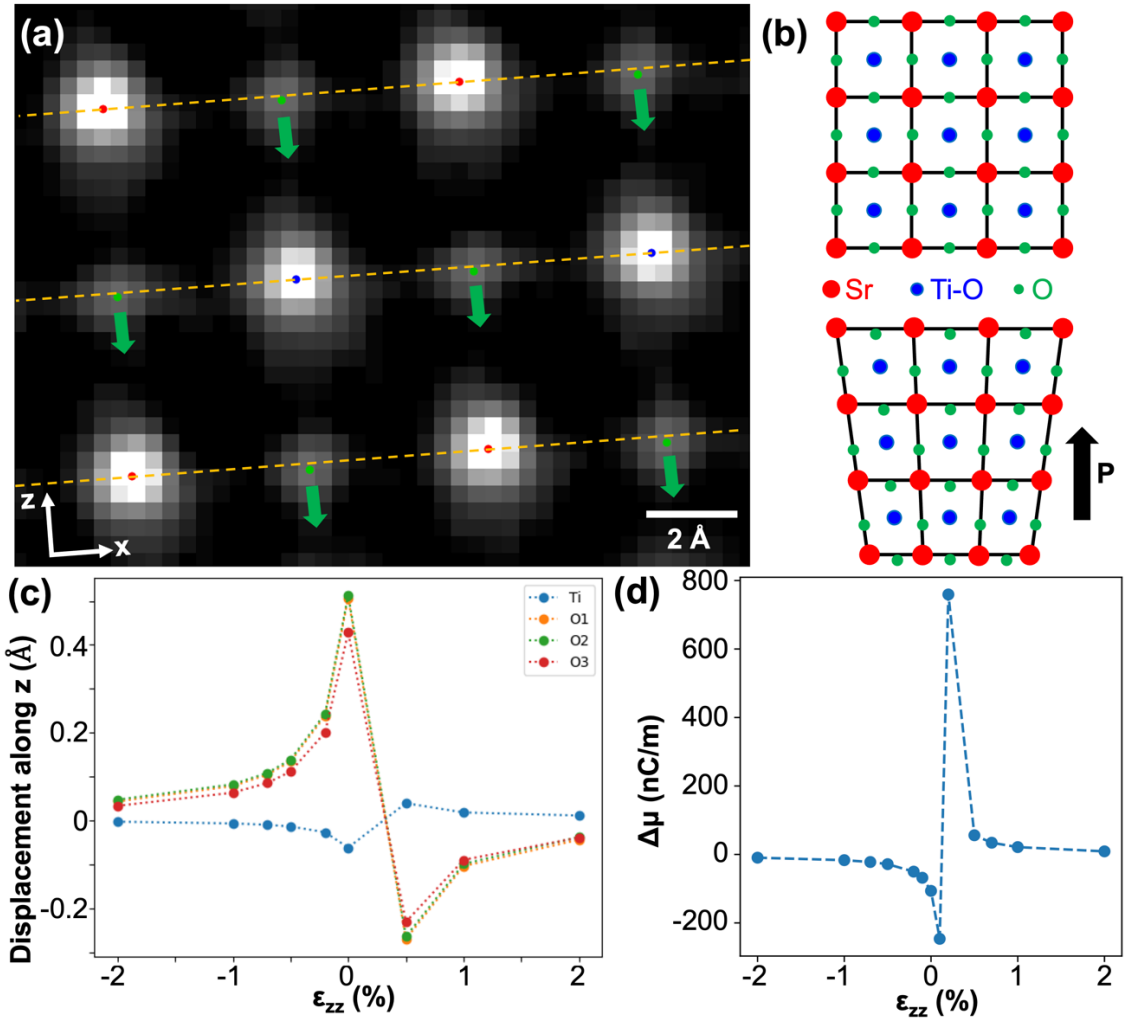

**Fig. 3 | *Microscopic origin of flexoelectricity and strain control of the flexoelectric coefficient.***
*a) Magnified experimental image of the region marked with the yellow box in Fig. 2(d), with the Sr, Ti-O and O atomic columns marked with red, blue and green dots respectively. The orange dotted line is drawn connecting the cation columns and shows the displacement of the oxygen atoms with respect to the atomic planes formed by the cation sublattices. (b) Schematic illustrating the undistorted perovskite structure (top) and the resulting distorted structure (bottom) on application of a strain gradient. The strain gradient as well as the off-centering of the oxygen atoms with respect to the cations is highly exaggerated for better visualization. (c) Displacements of the Ti and O atoms in the z-direction with respect to the Sr sublattice in response to a strain gradient of magnitude $10^7 m^{-1}$, calculated using DFPT and plotted as a function of the out-of-plane strain. The displacement of the O atoms is an order of magnitude larger than that of the Ti atom. The sign of the displacements switch in going from compressive to tensile strain. As the strain gradient direction remains unchanged, this strain-mediated switch in the sign of the flexoelectric polarization implies an effective switch in the sign of the flexoelectric coefficient from Equation 1. (d) Transverse flexoelectric coefficient (relative to its value when the strain and strain gradient-induced displacements result in zero polar distortion) plotted as a function of the out-of-plane strain, showing this switch in sign.*

An atomic scale characterization of the atomic displacements as demonstrated above with MEP provides definitive evidence about the direction of the ionic component of the polarization with respect to the strain gradient, allowing us to infer a positive sign of the flexoelectric coefficient. The measured positive sign for the flexoelectric tensor component of the form $\mu_{iijj}$ is consistent with experimental measurements on bulk $SrTiO_3$ by Zubko et al.[39], and Mizzi et al.[49], but opposite to that reported for bent $SrTiO_3$ membranes by Cai et al[22]. We address the origin of this discrepancy and support the observed oxygen displacement dominated mechanism using density functional perturbation theory[50,51] (DFPT) calculations.

As the presence of a strain gradient is inherently incompatible with periodic boundary conditions required for plane-wave calculations, standard DFPT approaches cannot be readily applied to tackle this problem. Instead, we adopt the approach introduced by Stengel[52–55], where flexoelectric responses are derived from a long-wave analysis of acoustic phonons. In the calculations, we use the form of a true bending-induced strain gradient of the form $\partial_z \varepsilon_{xx} - \upsilon\, \partial_z \varepsilon_{zz}$ (where $\upsilon$ is the Poisson ratio) which is the only way to truly have a static strain gradient from bending – details of our calculation are provided in Supplementary Text 1. We calculate the structural distortions in response to such a strain gradient and study how this response varies in the presence of a net uniaxial strain in either the in-plane (x) or out-of-plane (z) directions. In Fig. 3(c), we plot the calculated displacement of the Ti and O atoms (with Sr sublattice taken as reference) in response to a strain gradient magnitude of $10^7$ m$^{-1}$ as a function of uniaxial strain in the out-of-plane z-direction (see Supplementary Figure 9 for the case of in-plane strain). The magnitude of the oxygen atom displacements is consistently larger than that of Ti atoms for all strain values, in agreement with our experimental observations. Moreover, our calculations reveal a resonance-like response where the displacement magnitudes become substantially large, and across which the displacement direction switches, producing a corresponding switch in the direction of the flexoelectric polarization. As the direction of the strain gradient remains unchanged, this switch in the direction of the flexoelectric polarization implies a switch in the sign of the flexoelectric coefficient. This strain-induced switch in the sign of the transverse flexoelectric coefficient is shown in Fig. 3(d) where the theoretically calculated value of the coefficient is plotted as a function of the out-of-plane strain (the individual contributions to the flexoelectric response are shown in Supplementary Fig. 10). The flexoelectric coefficients in Fig. 3(d) are reported relative to their value when the strain and strain gradient-induced displacements result in zero polar distortion. We note that the absolute value is arbitrary due to the choice of the reference potential in the calculation.

The coupling between the strain and flexoelectric response is mediated by the hybridization of the polar optical phonon mode with the acoustic phonon modes on the application of strain, with the mechanism explained in Supplementary Text 2. At strain values where these bands strongly hybridize, the flexoelectric response is significantly amplified, exceeding the off-resonance response by several orders of magnitude. Given that these are zero-temperature calculations, the resonance peak will likely be damped by thermal vibrations at finite temperatures. However, it will still produce an enhanced response, albeit by a smaller factor.

Such strain-control of the flexoelectric coefficients in $SrTiO_3$ offers an alternative route to switching the flexoelectric polarization circumventing the need to reverse the direction of the strain gradient. We also note that the switching does not occur exactly at zero strain but rather at a small tensile (compressive) strain in the out-of-plane (in-plane) direction, that makes strain tuning across this transition point more favorable for practical applications. The observed trends are independent of the type of strain (uniaxial or biaxial) implemented in the calculation as shown in Supplementary Fig. 9. We also investigate the effect of surface piezoelectric contributions, electrostatic boundary conditions and lower symmetry space groups on the flexoelectric response with details provided in Supplementary Text 3.

**Electronic structure changes**

The large strain variations across the membrane as well as the polar distortion lead to deviations from the octahedral bonding environment of the Ti atoms, and will be reflected in the crystal field splitting energy (CFSE) of the Ti $3d$ orbitals. We use electron energy-loss spectroscopy (EELS) to probe these changes across the membrane. Figure 4(a) shows a series of EEL spectra in an energy range of 17 eV around the Ti $L_{2,3}$ edge collected from the top to the bottom of a bent region of the 15 nm thick membrane shown in Fig. 1. The Ti $L_2$ and $L_3$ transitions correspond to excitations from the spin-orbit coupling split $2p_{1/2}$ and $2p_{3/2}$ orbitals to the $3d$ orbitals respectively. In cubic perovskites, the octahedral geometry of the surrounding O atoms breaks the degeneracy of the $3d$ orbitals, which splits into $t_{2g}$ and $e_g$ orbitals and is reflected in the splitting of the $L_2$ and $L_3$ edges each into two sub-peaks. In the presence of strain or polarization, the degeneracies of the $t_{2g}$ and $e_g$ sublevels are also broken, with the transition energies from the $2p$ bands to the different $d$-bands systematically changing as a function of the strain and polarization[56]. The changes in relative ordering as well as the energies of the different $d$-orbitals are reflected in the shape and position of the EELS edges. These details are limited by the core-hole lifetime broadening and the finite energy resolution (~1 eV) of our probe, and we track the overall shift in the splittings between the $t_{2g}$ and $e_g$ manifolds.

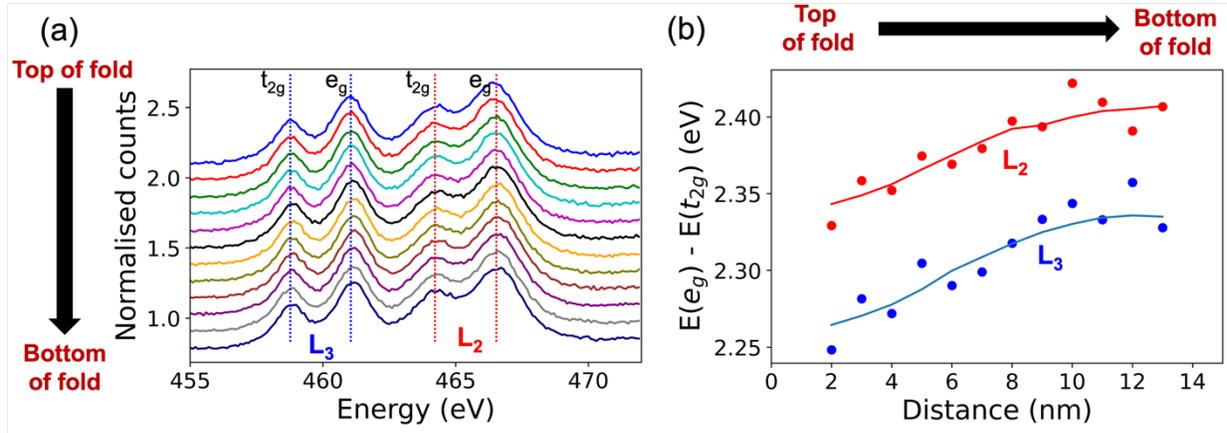

*Fig. 4 | Tracking electronic structure changes with EELS. a) Waterfall plot of experimental EELS spectra taken across the fold from the top to bottom, showing the Ti $L_{2,3}$ edge. Each displayed spectra is a sum of 860 individual spectra collected over an 8.6 nm x 1 nm region. The spectra at the top and bottom surfaces are omitted due to additional effects from surface reconstructions. Each spectra is fit with a sum of four Lorentzians to calculate the peak positions of the $t_{2g}$ and $e_g$ peaks of the $L_2$ and $L_3$ edges. b) Difference between the energies of the $e_g$ and $t_{2g}$ peaks for the $L_2$ and $L_3$ edges, where the solid line is a fit with a Savitzky-Golay smoothing filter. This difference scales with the crystal field splitting energy and shows a gradual increase from the top to the bottom of the fold (opposite to the strain gradient direction).*

We use Lorentzian functions to fit all the 4 main peaks in each spectrum in Fig. 4(a) - the energy difference (ΔE) between the $t_{2g}$ and $e_g$ peaks in both the $L_2$ and $L_3$ edges is directly related, although not equal to the crystal field splitting energy (CFSE) in the system[57]. The energy difference between the $t_{2g}$ and $e_g$ peaks as a function of distance across the membrane is shown in Fig. 4(b). We observe a near-monotonic trend, with the magnitude of the CFSE increasing as we move from the top to the bottom of the fold, with an overall variation as high as 70 meV. As we move from the top (tensile strain) to the bottom (compressive strain) of the fold, the Ti-O bond length gradually decreases, resulting in a larger magnitude of the crystal field splitting energy.

Understanding the exact origins of the change in crystal field splitting energy requires a deeper look into the microscopic behavior of the Ti 3*d* band splitting in response to the strain gradient and the flexoelectric polarization and is discussed in Supplementary Text 4.

With increasing miniaturization of modern-day devices, large strain gradients can develop locally or globally, affecting electronic, dielectric and transport properties[9,58–62]. A fundamental understanding of the flexoelectric effect is essential to not only engineer devices taking advantage of the large magnitude of the effect at small length scales, but also to accurately model, predict and design the properties of all kinds of nanoscale devices. Our findings clarify the significant discrepancies in the values of the flexoelectric polarization reported in literature, highlighting its sensitivity to factors such as the strain state, soft optical phonon modes, and electrostatic boundary conditions. We show that the magnitude and sign of the flexoelectric coefficient is coupled to the strain state of $SrTiO_3$ membrane, indicating switchability by tuning the strain state without the need to change the strain gradient direction. With the experimental capabilities to engineer precise geometries and study their properties at the atomic scale as demonstrated here, flexoelectricity offers promise for the development of the next generation of electro-mechanical devices.

## Methods

### Nanopillar array fabrication

We fabricate the nanopillar structure via electron-beam lithography (JEOL 6300). First, we spin-coat the P20 primer and the s1813 photoresist subsequently onto a $SiO_2$ (500nm)/p-Si (525 m) wafer. After coating, the wafer is baked at 115°C for 1 minute and then developed in AZ 726

MIF. We use the GCA 6300 DSW 5X g-line wafer stepper to define alignment marks on the wafer, followed by buffer oxide wet etching (BOE 6:1) and inductively coupled plasma etching (Oxford Cobra Hbr etcher). Next, we spin-coat the SURPASS 4000 adhesion layer and the negative M-aN 2403 resist, followed by a soft bake at 90°C for 1 minute. After aligned exposure with electron-beam lithography, the wafer is developed in AZ 726 MIF for 70 seconds. Cylindrical nanopillars are then etched into the $SiO_2$ via reactive ion etching. The resultant nanopillars have a 150 nm diameter and a 300 nm height.

**Growth and transfer of SrTiO₃ free-standing membrane**

Epitaxial bilayer films of 16 nm thick $Sr_2CaAl_2O_6$ and {6, 15} nm thick $SrTiO_3$ were grown on a $SrTiO_3$ (001) oriented substrate (Shinkosha Ltd, Japan) via pulsed laser deposition by ablating targets of $Sr_2CaAl_2O_6$ and $SrTiO_3$ at a laser fluence of 1.25 $Jcm^{-2}$ and 0.45 $Jcm^{-2}$ respectively with an oxygen partial pressure of $5\times10^{-6}$ Torr. The growth thickness was monitored during deposition using *in situ* RHEED. For release of the $SrTiO_3$ membrane, the sample was spun coated with 600 nm of PMMA (950 K, 9% in Anisole). The sample was then kept in water for ~3 days for release of the membrane. The membrane with the supporting PMMA layer was floated to the surface of water and scooped with the silicon wafer with the nanopillar array. The sample was then transferred onto a hotplate at 80°C for 15 mins. The PMMA was subsequently removed using dissolution in warm acetone at 40°C.

**STEM characterization**

The cross-sectional TEM samples were prepared using a Thermo Fisher Scientific Helios G4 X focused-ion beam (FIB). Electron beam assisted carbon deposition was done on either side of the wrinkle to offer mechanical stability before the standard lift-out procedure using a gallium ion beam. HAADF images were acquired using an aberration corrected FEI Titan Themis X-FEG microscope with a semi-convergence angle of 21.4 mrad or on a Thermo Fisher Spectra X-CFEG with a semi-convergence angle of 30 mrad, both operated at 300 kV. The nanobeam electron diffraction patterns to show Friedel symmetry breaking were extracted from 4D STEM datasets acquired on a EMPAD-G2[63] detector installed on the FEI Titan Themis with a semi-convergence angle of 2.14 mrad.

**Strain mapping from HAADF images**

The positions of the Sr and Ti atoms in the HAADF images were measured using 2D Gaussian fitting, making use of the DAOStarFinder[64] function in the photutils[65] package as well as the Atomap[66] library in Python. For each Sr atom, the distance to the 2 nearest neighbors in both pseudo-cubic directions were calculated and averaged to get the local lattice constants in the two directions. The pixel size calibration was done using images of the silicon substrate.

**Multislice Electron Ptychography (MEP)**

The 4D-STEM datasets required for the MEP reconstructions were acquired on the EMPAD[67] detector (128 x 128 pixels) using an aberration corrected Thermo Fisher Spectra 300 X-CFEG microscope with a semi-convergence angle of 30 mrad and a total beam current of 60 pA. The EMPAD detector was run at a frame rate of 1 kHz with an outer collection angle of 53.3 mrad. The datasets were acquired with a field of view of around 10.6 nm x 10.6 nm and a probe overfocus of 10 nm.

The reconstructions were done using PtyRAD, an in-house Python based ptychographic package (currently being prepared for public release) based on automatic differentiation. We first ran an initial reconstruction without a tilt propagator and used the interlayer shifts between the different slices in the resulting object to generate an initial guess for the probe-position dependent tilt. This initial guess of the tilts was then optimized in a second reconstruction where a tilt propagator was used to account for spatial variations in sample tilt. We use a slice thickness of 1 nm and 6 probe modes to account for partial coherence of the beam[68,69].

**Electron Energy Loss Spectroscopy (EELS)**

The EEL spectra were recorded with a 965 GIF Quantum ER spectrometer and a Gatan K2 Summit direct electron detector operated in the pulse-counting mode, installed on an aberration corrected FEI Titan Themis microscope. The energy resolution measured using the full width half-maximum is ~1 eV. Spectra were collected in the energy range of 330-700 eV using a dispersion of 0.1 eV per channel. The microscope was operated at 300 kV with a probe semi-convergence angle of 21.4 mrad. The spectra were collected near the top of the wrinkle, with a scan step of 0.1 nm and a scan area of width 8.6 nm and height of 19.8 nm spanning the 15 nm thick membrane.

In order to get sufficient signal to noise ratio for mapping out the small shifts in the peaks, a beam current of 60 pA and a dwell time of 9.8 ms per pixel was used. Since we are interested in the spectral variations parallel to the strain gradient, all the spectra perpendicular to the strain gradient direction are averaged. Further, the mean of all spectra in a distance range of 1 nm parallel to the strain gradient direction is taken to produce each plotted spectra in Fig. 3(a).

Hence, each spectrum in Fig. 3(a) is calculated as an average of 860 individual spectra spanning 1 nm parallel to and 8.6 nm perpendicular to the strain gradient direction.

**DFPT calculations**

Calculations were performed using the methodology for computing flexoelectric coefficients in the ABINIT[70,71] code. We used a theoretically determined lattice parameter of 3.857 Å, the LDA exchange-correlation functional, optimized norm-conserving Vanderbilt pseudopotentials[72] from PSEUDODOJO[73], a 10×10×10 mesh of k-points to sample reciprocal space, and a plane-wave cutoff of 80 Ha.

**Acknowledgements**

H.K., V.H., J.C., K.J.C., Y.-T.S., G.D.F., H.Y.H., and D.A.M. acknowledge funding from the Department of Defense, Air Force Office of Scientific Research under award FA9550-18-1-0480. The work at Stanford/SLAC was supported by the U. S. Department of Energy, Office of Basic Energy Sciences, Division of Materials Sciences and Engineering (Contract No. DE-AC02-76SF00515). C.-H.L. would like to thank the support from the Eric and Wendy Schmidt AI in Science Postdoctoral Fellowship, a program of Schmidt Sciences, LLC. This work made use of the electron microscopy facility of the Platform for the Accelerated Realization, Analysis, and Discovery of Interface Materials (PARADIM), which is supported by the National Science Foundation under Cooperative Agreement No. DMR-2039380 and Cornell Center for Materials Research shared instrumentation facility with Helios FIB supported by NSF (DMR-1539918) and FEI Titan Themis 300 acquired through NSF-MRI-1429155. The authors also thank John

Grazul, Mariena Silvestry Ramos and Malcolm Thomas for technical support and maintenance of the electron microscopy facilities.

## Author Contributions

The research plan was formulated by H.K., V.H., G.D.F., H.Y.H. and D.A.M. H.K. performed STEM and MEP characterization under the supervision of Y.-T.S. and D.A.M. PLD synthesis, liftoff, and transfer of the oxide membranes onto the nanopillar array were done by V.H., K.J.C. and Y.L. under the supervision of H.Y.H. J.C. fabricated the nanopillar array under the supervision of G.D.F. C.-H.L. developed routines for position-dependent tilt correction and ran ptychographic reconstructions. D.Y. and H.K. performed SEM imaging of the nanopillar arrays and wrinkles. All the theory calculations were done by C.D.

## Competing interests

Cornell University (D.A.M.) has licensed the EMPAD hardware to Thermo Fisher Scientific. The remaining authors declare no competing interests.

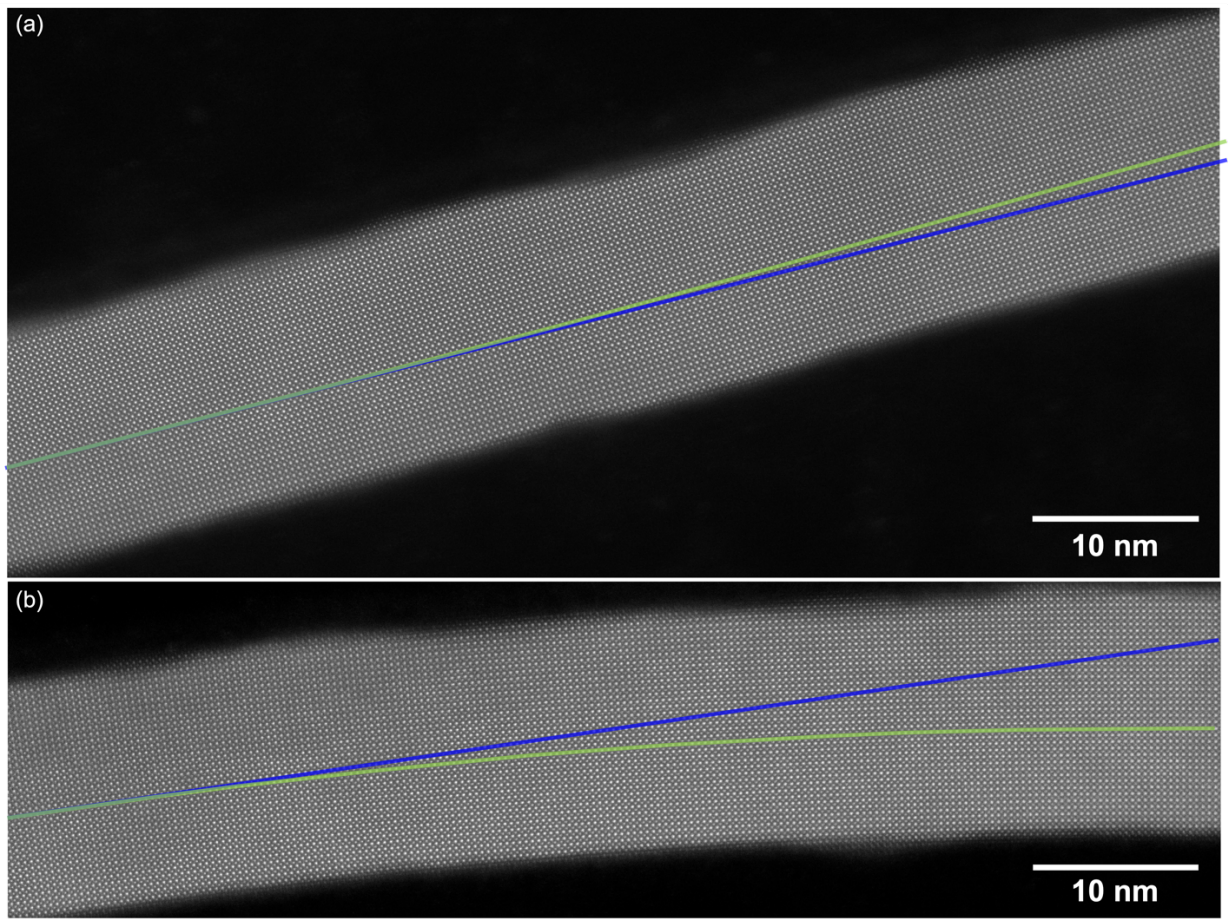

**Supplementary Fig. 1 | Increased curvature of atomic planes near the wrinkle apex.** HAADF-STEM image of the wrinkle (a) away from the apex and (b) close to the apex. Both green and blue lines start at the same atomic planes, with the blue line following a straight-line path while the green line follows the curvature of the atomic planes. The curvature of the atomic planes is much larger near the wrinkle apex in (b) compared to away from the apex in (a).

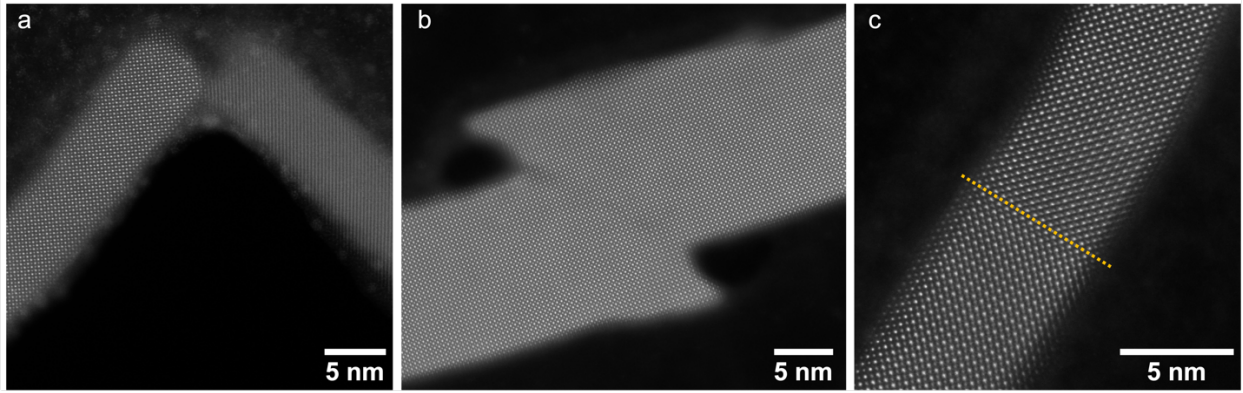

**Supplementary Fig. 2 | Alternative strain relaxation mechanisms. (a)** Rupture at the wrinkle apex due to the local strain exceeding the thickness-dependent fracture strain of the membrane. **(b, c)** Relaxation of the shear strain at the point of contraflexure by a (b) slip defect or (c) rearrangement of atomic planes reminiscent of a deformation twin boundary. Prior work on bending, slipping and twinning in 2D materials[1,2] can potentially be extended to study the strain relaxation mechanisms in oxide membranes.

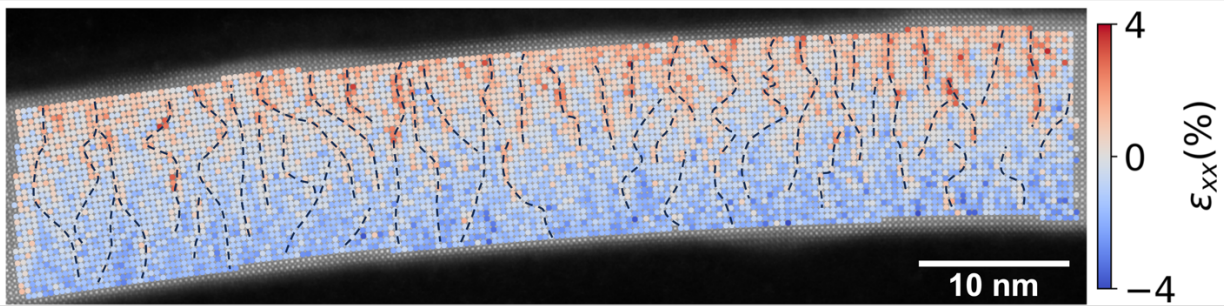

**Supplementary Fig. 3 | Inhomogeneity in strain profile.** Finger-like structures in the strain map showing the local inhomogeneity in the strain distribution. Some of the more prominent finger-like structures that show higher (lower) tensile (compressive) strain compared to its neighboring areas are labelled with black dotted lines.

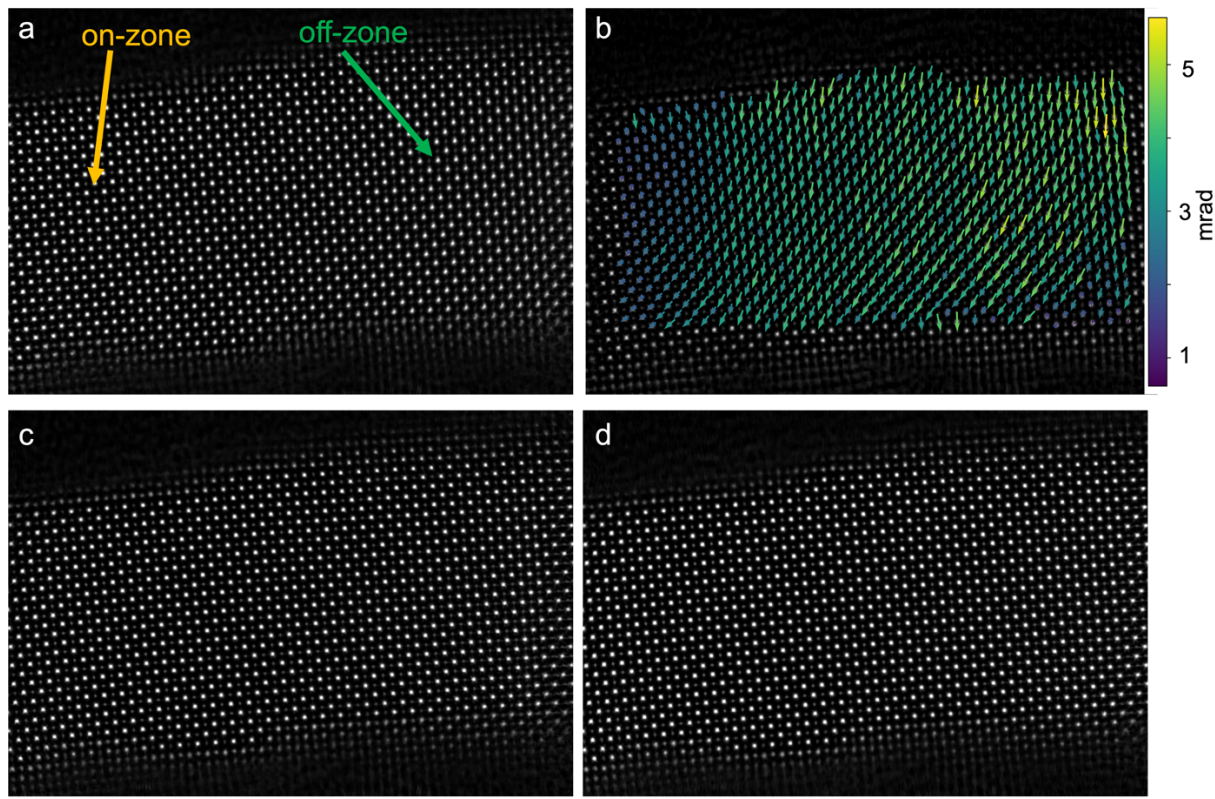

**Supplementary Fig. 4 | Misorientation correction for MEP reconstruction of the wrinkle apex in the 6 nm thick membrane.** (a) MEP image (summed along the depth direction) reconstructed without the tilt propagator. Due to sample mistilts in the off-zone region, atoms get elongated in projection causing systematic errors in their tracked positions. (b) Quiver plot showing the spatially varying sample tilt calculated from the lateral shift in the atomic positions with depth. A smooth envelope function is used to fit the measured tilts and compute the tilt value at each probe position, used as initial guesses for the tilt propagator. (c, d) MEP images from reconstructions with the tilt propagator, (c) keeping the tilt values fixed, and (d) allowing the tilt values to update. The reconstructed sample region contains an on-zone area with minimal mistilt, labelled in (a), which is not affected by the misorientation correction. The measured displacements in the on-zone region are consistent with the overall trend, serving as an additional validation and confirming that any minor systematic errors from the correction do not impact our conclusions.

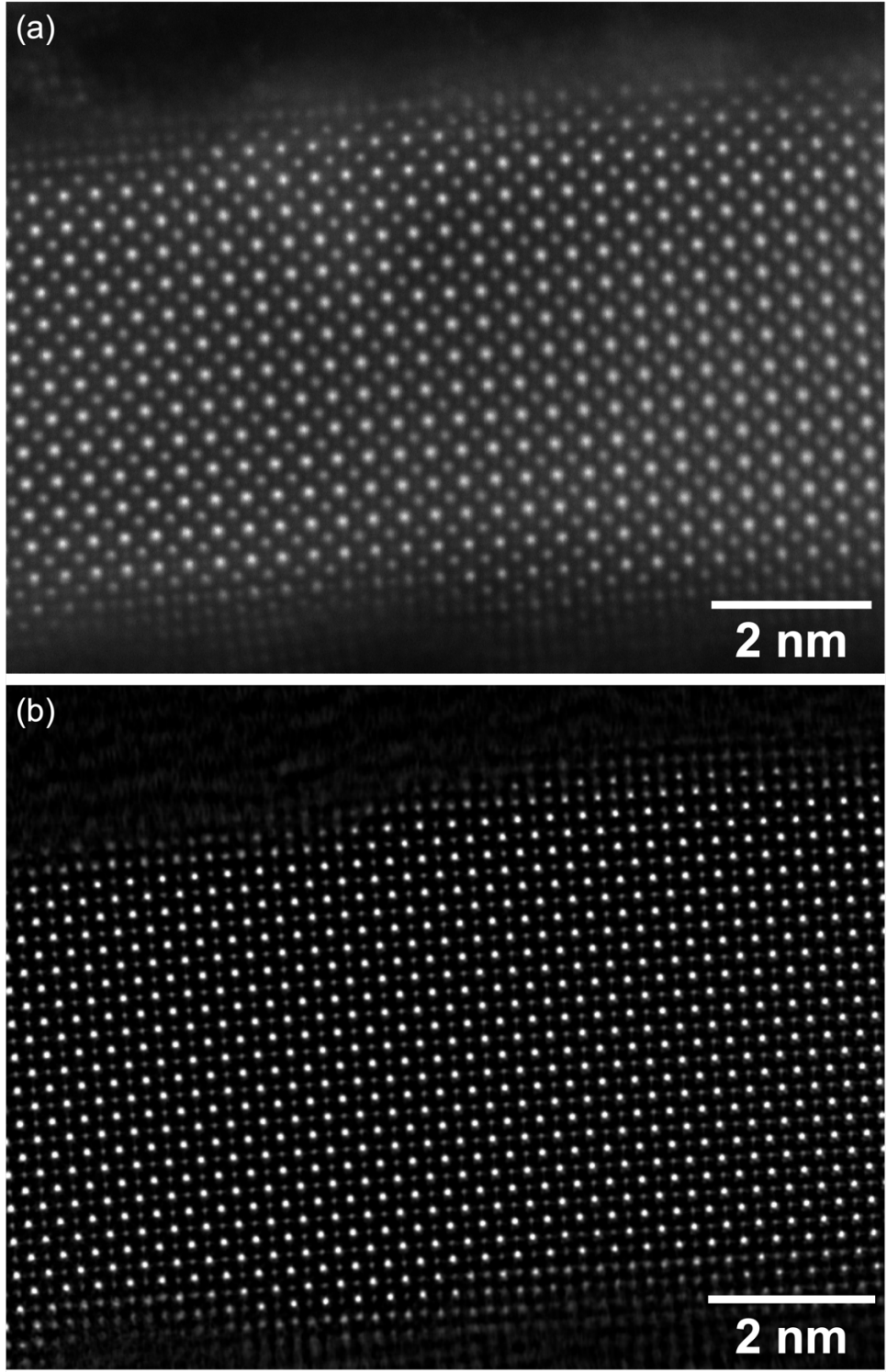

**Supplementary Fig. 5 | Comparison of HAADF and MEP images of the wrinkle apex in the 6 nm thick membrane.** The (a) HAADF image is insensitive to the light oxygen atom positions, which are crucial in getting an accurate picture of the polarization. (b) MEP generates a higher resolution image in comparison, with all the light oxygen atoms reliably resolved.

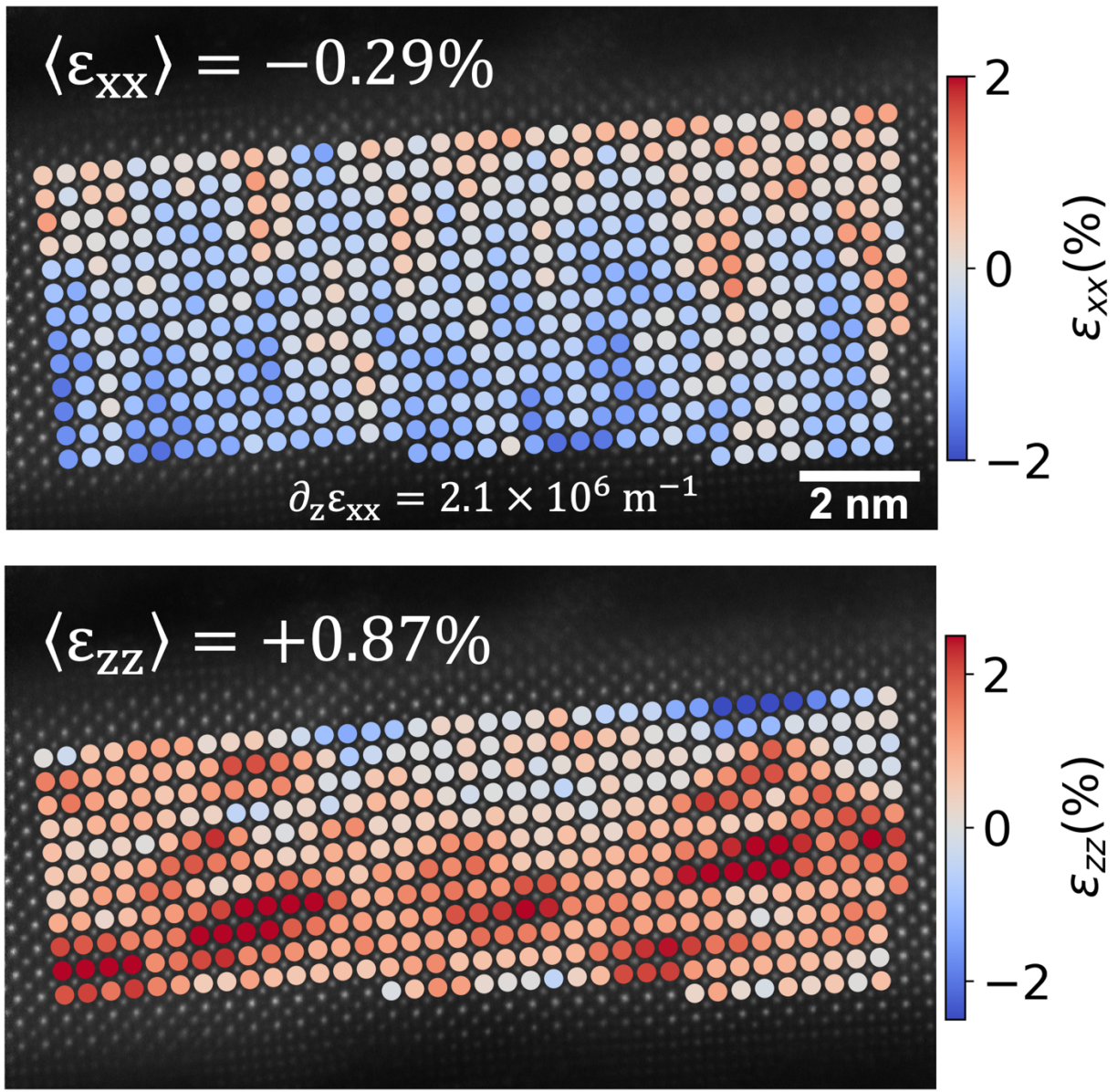

**Supplementary Fig. 6 | Strain maps calculated for the 6 nm thick membrane.** Strain maps showing the variations in the local lattice parameter along the **(a)** in-plane direction and **(b)** out-of-plane direction. The maps reveal the presence of a transverse strain gradient like that shown in Fig, 1(d) for the 15 nm thick membrane. Although less obvious, the strain map for the out-of-plane lattice parameter also displays an overall longitudinal strain gradient, likely associated with strain relaxation governed by the Poisson ratio. A brief discussion is provided in Supplementary Text 1.

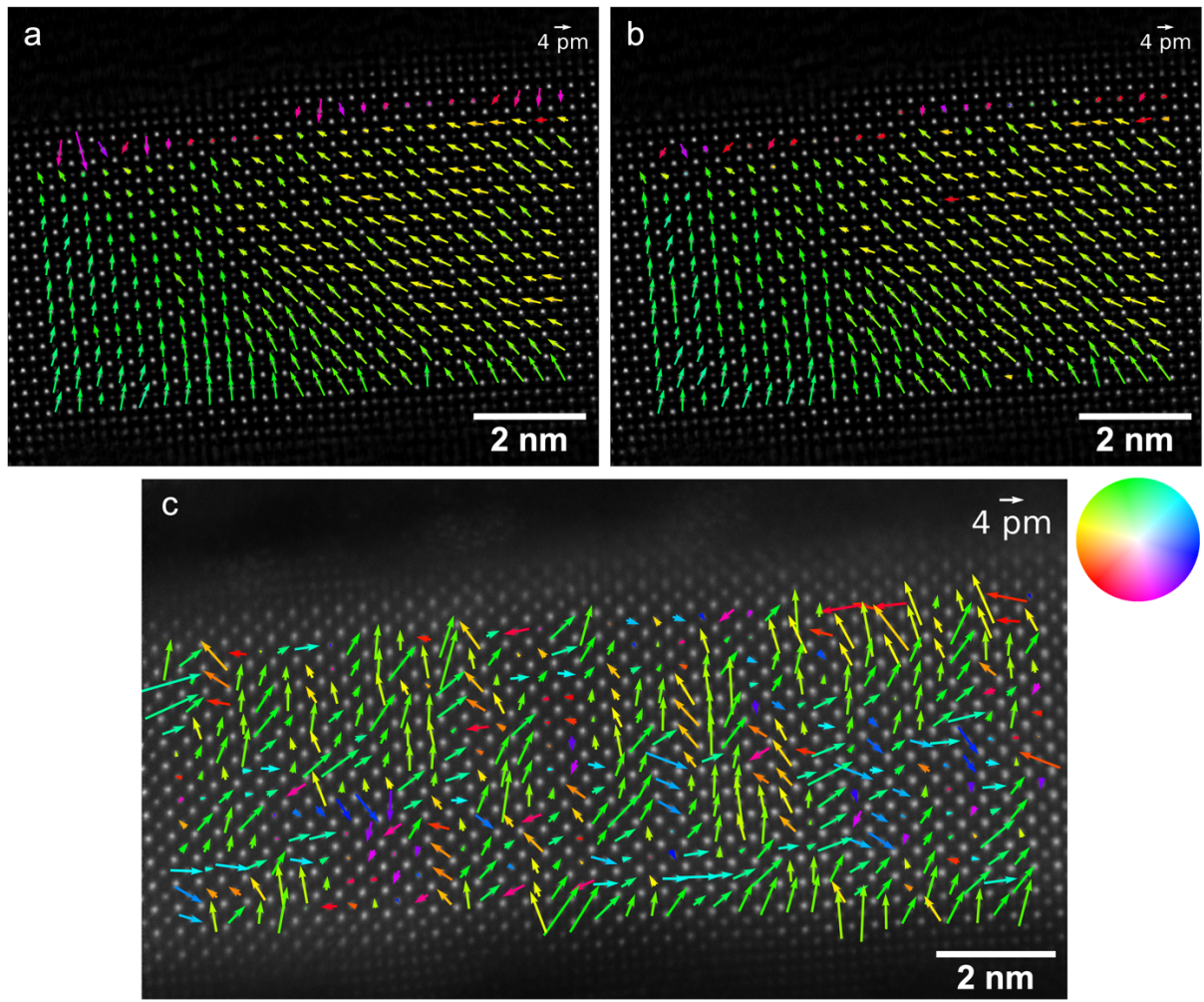

**Supplementary Fig. 7 | Atomic displacement maps.** Vector map of the displacement of the (a) centroid of Sr atoms and (b) Ti atom with respect to the centroid of the O atoms for each unit cell calculated from and overlaid on the MEP image. (c) Vector map of the cation-cation displacements (displacement of the Ti atom with respect to the centroid of the Sr atoms) that is often used as a proxy for the polarization calculated from the HAADF image. As our measurements from the MEP image indicated minimal cation-cation displacements, this map generated from the HAADF image showing randomly oriented vectors is just noise. Fig. 2(d) is the same as subfigure (b), while Fig. 2(c) shows a cropped region from subfigure (c). They are reproduced here for clarity.

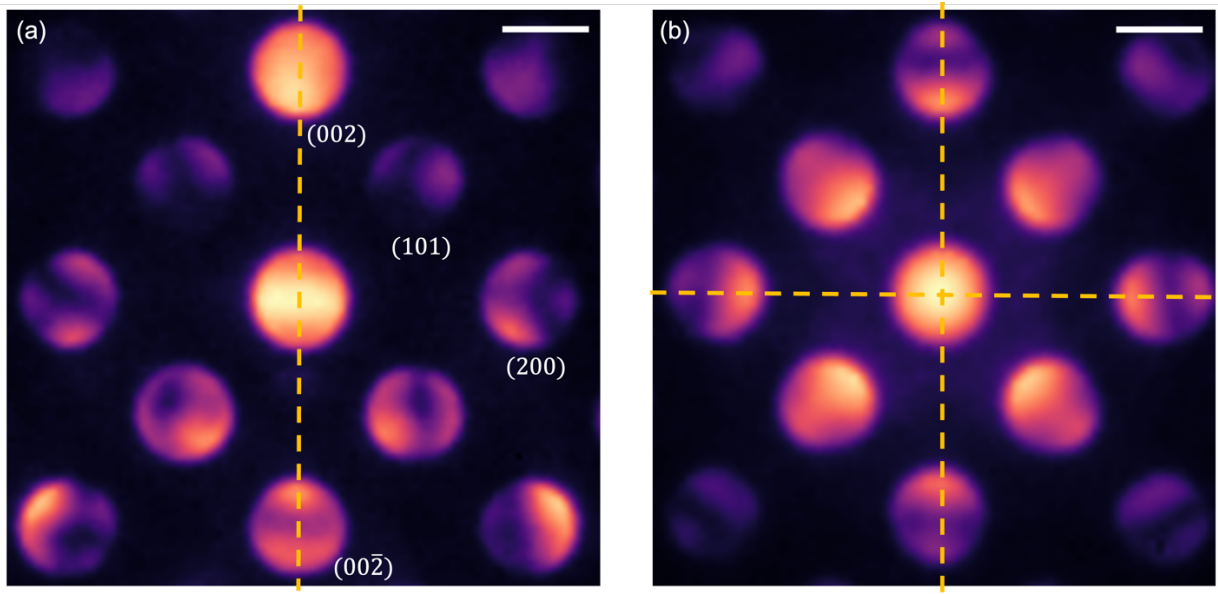

**Supplementary Fig. 8 | Detecting polarization from breaking of Friedel symmetry in CBED patterns.** CBED patterns taken from the STO membrane (a) near the wrinkle apex and, (b) at a flat region of the 15 nm thick membrane. In (a), the (002) and (00$\bar{2}$) Friedel pairs of diffraction spots show a clear asymmetry due to the polarization in the (001) direction. The mirror symmetry with normal vector (001) along the polarization direction is broken, while the mirror symmetry with normal vector (100) perpendicular to the polarization direction remains almost intact. The small deviations from perfect mirror symmetry in the latter case can be attributed to flexoelectric contributions from the presence of small gradients in the out-of-plane lattice constant. In (b), both mirror symmetries are preserved as there is little to no polarization in the flat region where atomic planes do not have any curvature. Scalebar is 2 nm$^{-1}$.

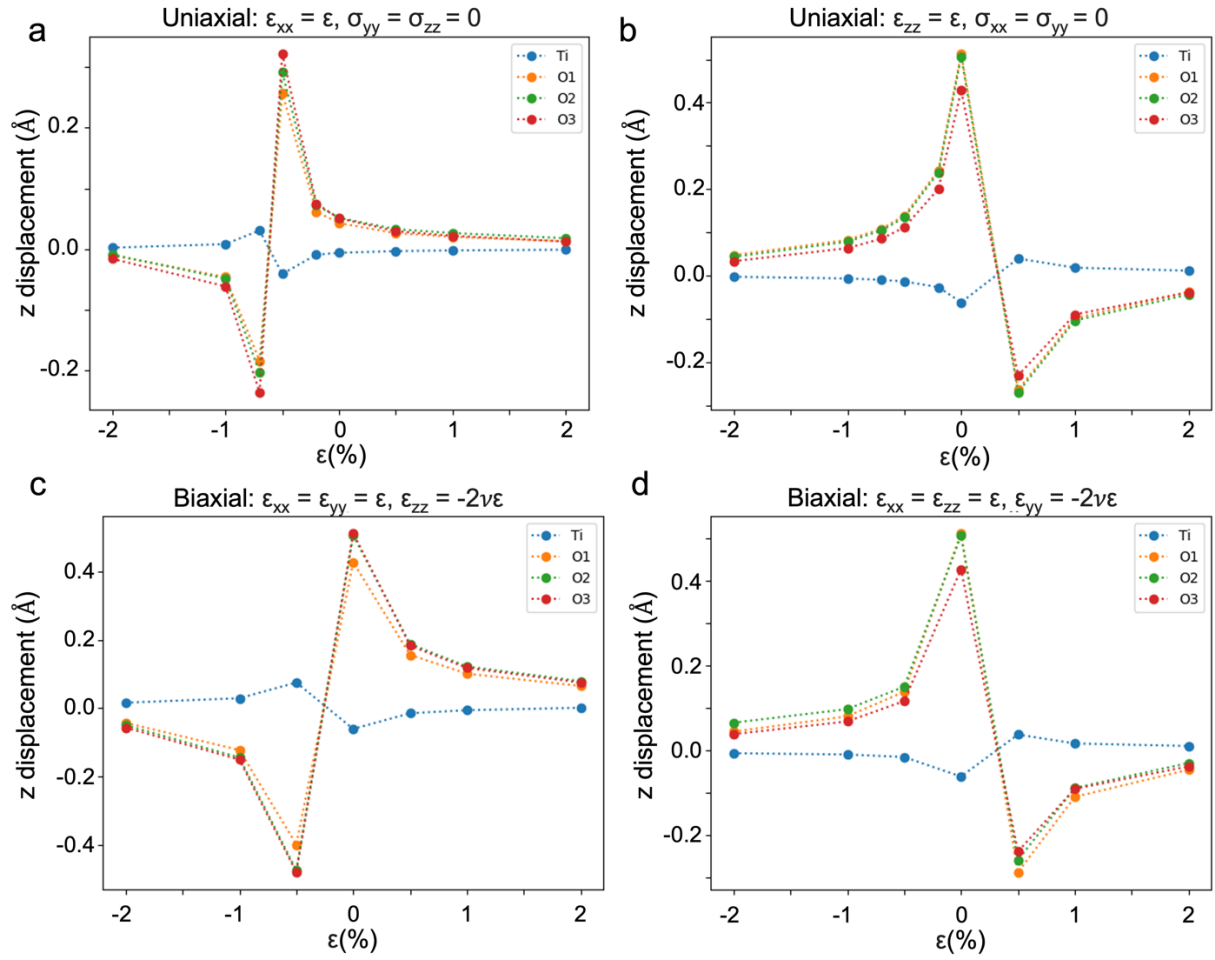

**Supplementary Fig. 9 | Trends in atomic displacements as a function of biaxial and uniaxial strain.** For an effective strain gradient of the type $\partial_z \varepsilon_{yy} - \nu\, \partial_z \varepsilon_{zz}$ with magnitude 0.001 Å$^{-1}$, the atomic displacements of the Ti and O atoms are plotted as a function of (a, b) uniaxial and (c, d) biaxial strain. Irrespective of the type of strain, the Ti displacement is consistently lower in magnitude compared to that of the oxygen atoms. The sign of the bulk contribution to the flexoelectric coefficient (opposite to the sign of the oxygen displacement) is positive for a large out-of-plane (in-plane) tensile (compressive) strain, and negative for a large out-of-plane (in-plane) compressive (tensile) strain.

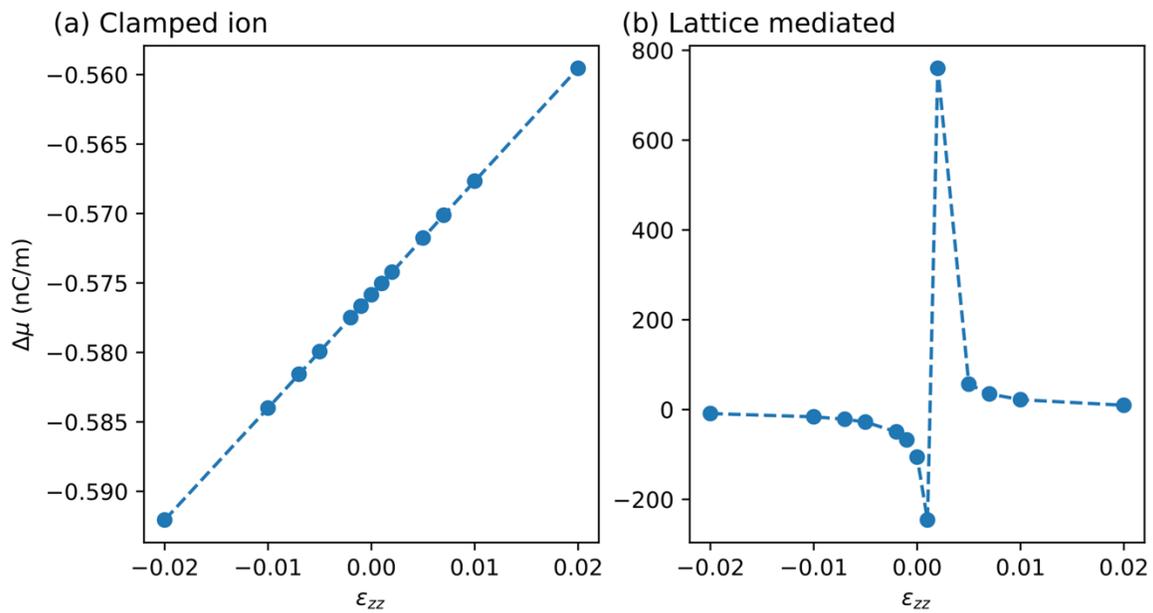

**Supplementary Fig. 10 | Individual contributions to the flexoelectric coefficient.** (a) Clamped-ion and (b) lattice-mediated contributions to the flexoelectric coefficient. The clamped-ion contribution arising from the valence electrons is negligible in comparison to the lattice-mediated response. As in Fig. 3(d), the flexoelectric coefficients are reported relative to their value when the strain and strain gradient-induced displacements result in net zero polar distortion.

# Supplementary Text 1 – Theoretical background

In the discussion below, we follow the notation used by Stengel[3] (this is distinct from the notation we used in the main text). The flexoelectric (FxE) coefficients in type-II notation, i.e., with respect to symmetrized strains as opposed to unsymmetrized displacements, is given by:

$$\mu_{\alpha\lambda,\beta\gamma} = \frac{\partial P_\alpha}{\partial \varepsilon_{\beta\gamma,\lambda}} \tag{1}$$

where $P_\alpha$ is the electric polarization in direction α and $\varepsilon_{\beta\gamma,\lambda}$ is the strain gradient given by

$$\varepsilon_{\beta\gamma,\lambda} = \frac{\partial \varepsilon_{\beta\gamma}}{\partial r_\lambda} \tag{2}$$

where $\varepsilon_{\beta\gamma}$ is symmetric with respect to swapping β and γ, and therefore so in $\mu_{\alpha\lambda,\beta\gamma}$. Note that we will always be treating linear response taken around zero strain gradient. We can write the flexoelectric coefficients as:

$$\mu_{\alpha\lambda,\beta\gamma} = \bar{\mu}_{\alpha\lambda,\beta\gamma} - \bar{P}^{(1,\lambda)}_{\alpha,\kappa\rho} \Gamma^{\kappa}_{\rho\beta\gamma} + \frac{1}{\Omega} Z^*_{\kappa,\alpha\rho} L^{\kappa}_{\rho\lambda,\beta\gamma} \tag{3}$$

Note that repeated indices are summed over. The first term in Eq. (3) is the clamped-ion response. It is not of interest to us since it does not involve relaxation of the lattice; it can be thought of as the response of just the valence electrons to the symmetry breaking caused by the strain gradient. The second term has two parts: $\bar{P}^{(1,\lambda)}_{\alpha,\kappa\rho}$ can be thought of as an atom-resolved (κ corresponds to the sublattice) clamped-ion piezoelectric coefficient with respect to a strain $\varepsilon_{\rho\lambda}$. $\Gamma^{\kappa}_{\rho\beta\gamma}$ is the piezoelectric internal strain tensor, which corresponds to the internal relaxation of sublattice κ in direction ρ resulting from an $\varepsilon_{\beta\gamma}$ strain. The last term also has two parts (Ω is simply the unit cell volume). $Z^*_{\kappa,\alpha\rho}$ is the element of the Born effective charge tensor for sublattice κ. $L^{\kappa}_{\rho\lambda,\beta\gamma}$ is the flexoelectric internal strain tensor, which gives the internal relaxation of sublattice κ in direction ρ resulting from a strain gradient $\varepsilon_{\beta\gamma,\lambda}$.

Now consider a material under a constant strain gradient. We will have two contributions to the internal relaxations resulting from this gradient to linear order. First, we have an internal relaxation directly caused by the strain gradient, which is parameterized by **L**. This is constant throughout the material (assuming the strain gradient is constant everywhere). Second, considering a unit cell at a given point in the material, it will also be under some amount of uniform strain, resulting in internal relaxations parameterized by **Γ**. Of course, this internal relaxation will not be uniform throughout the material, since the strain is not uniform. In the case of cubic $SrTiO_3$, **Γ** = 0 by symmetry, and thus there is no effect from this contribution.

We must account for one more consideration before performing calculations. That is, if we consider the strain gradient to be statically applied, we must have mechanical equilibrium everywhere in the sample. A rough conceptual picture is the following - consider a longitudinal strain gradient $\varepsilon_{zz,z}$ which looks like the displacement pattern of a long-wavelength longitudinal acoustic phonon. There is no way we can impose this strain gradient statically. If we use atomic tweezers to place the atoms in this gradient and then let them go, the strain gradient will relax away, resulting in (possibly) a uniform strain in the z direction. Mathematically, mechanical equilibrium can be expressed as:

$$\sum_{\beta\gamma\lambda} C_{\alpha\lambda,\beta\gamma} \varepsilon_{\beta\gamma,\lambda}(r) = 0 \qquad (4)$$

for every point r in the sample, where **C** is the elastic tensor. This basically means that a single component of the strain gradient tensor cannot be sustained; a static deformation field involves two or more. Following Stengel[4], we consider an "effective" bending strain gradient of the form

$\varepsilon_{\text{eff},z} = \varepsilon_{xx,z} - \nu\varepsilon_{zz,z}$, where z is normal to the surface, and $\nu = \frac{C_{zz,xx}}{C_{zz,zz}}$ is the Poisson ratio (for SrTiO$_3$, $\nu = 0.289$).

## Supplementary Text 2 – Mechanism of strain-induced switching of the flexoelectric polarization

To determine the reason for the significant dependence of the strain-gradient-induced polar distortions on the strain in the crystal we calculated the phonons in SrTiO$_3$ under uniaxial strain as shown in Supplementary Fig. 11 (we used the same computational parameters as the flexoelectricity calculations). We can clearly see that one of the branches of the transverse optical mode at the zone center (highlighted in red in Supplementary Fig. 11) is softened by either tensile or compressive strain. This mode exactly corresponds to the polar distortion resulting from displacements of O versus Ti atoms. At the strain where there is an abrupt change in the sign of the polar distortion (~ -0.005, see Fig. 3 in the main text), we see that this mode is almost degenerate with the acoustic modes (blue in Supplementary Fig. 11). Thus, around this strain, there is a large coupling between the strain gradient (which can be described as the second-order response to an acoustic phonon perturbation) and the polar distortion.

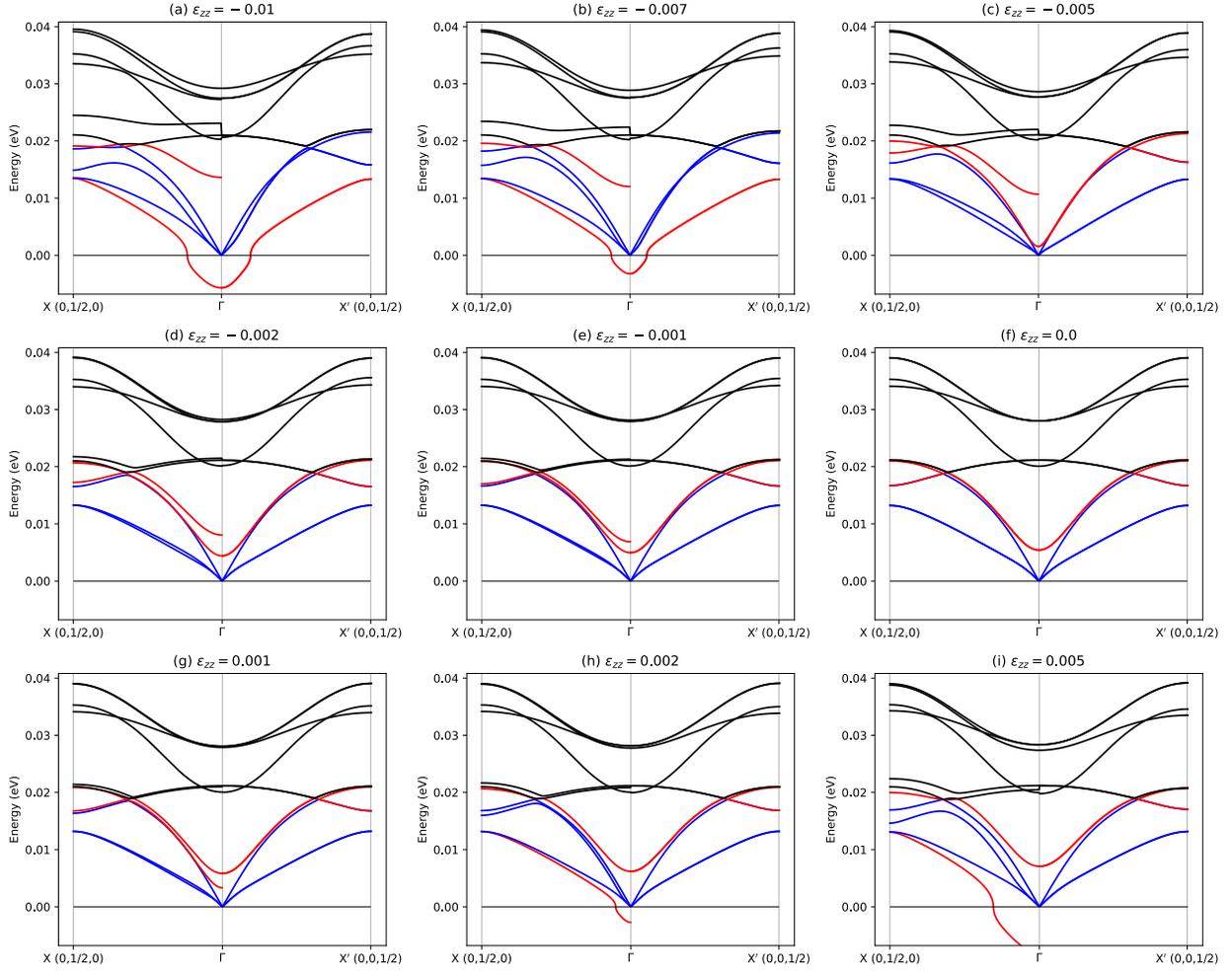

**Supplementary Fig. 11** | Calculated phonon dispersions with respect to uniaxial strain in SrTiO$_3$.

**Supplementary Text 3 - Theoretical exploration of the effect of other potential parameters on the flexoelectric polarization:**

### i)  Effects of lower symmetry

We consider the possibility of symmetry lowering distortions which may be present because of the strain/strain gradient. We can do this explicitly by considering PbTiO$_3$, for which *P4mm* is a metastable structure. Then, $\mathbf{\Gamma} \neq 0$ (where $\mathbf{\Gamma}$ is the piezoelectric internal strain tensor defined in

Equation 3 in the Methods section), and we must consider the atomic relaxation resulting from local strain. At each point in the film, the effective bending will result in $\varepsilon(\mathbf{r}) = \varepsilon_{xx}(\mathbf{r}) - \nu\, \varepsilon_{zz}(\mathbf{r})$, where ν is the Poisson ratio. If we take as our origin the point of zero strain in the middle of the film, we will have $\varepsilon_{xx} > 0$ ($\varepsilon_{xx} < 0$) for z > 0 (z < 0) and $\varepsilon_{zz} > 0$ ($\varepsilon_{zz} < 0$) for z < 0 (z > 0). The results are shown in Table I. First, we see that we would expect to see different signs of the displacements depending on the z position, which is not seen experimentally. Also, we see that the difference in Ti and O displacements that we saw for the cubic case (Fig. 3c) is no longer present, again in contradiction with experiment. Again, we attribute this to hardening of the soft polar mode at Γ, which has condensed in the *P4mm* structure. Though this calculation is for PbTiO$_3$, we expect the same to occur if *P4mm* SrTiO$_3$ is stabilized.

|  | Ti | O1 | O2 | O3 |
|---|---|---|---|---|
| $\varepsilon_{xx,z}$ | 0.017 | 0.039 | 0.037 | 0.034 |
| $\varepsilon_{zz,z}$ | 0.018 | 0.039 | 0.039 | 0.036 |
| $\varepsilon_{eff,z}$ | 0.01 | 0.023 | 0.021 | 0.02 |
| $\varepsilon_{eff} = \pm 0.02$ | $\mp 0.017$ | $\pm 0.011$ | $\mp 0.012$ | $\mp 0.002$ |
| Total ($\pm 0.02$) | $\pm 0.027$ | $\pm 0.012$ | $\pm 0.033$ | $\pm 0.022$ |

**TABLE I.** Atomic displacements in Å in the z direction with respect to the Pb atoms in *P4mm* PbTiO$_3$ for transverse, longitudinal, and effective bending strain gradients with magnitude 0.0007 Å$^{-1}$ and an effective strain that varies from -2% to +2%.

ii) **Surface piezoelectric contributions**

Strain at the surface of the film will always cause a dipole because inversion symmetry is broken at the surface. Since there is a strain gradient, and thus opposite strain on opposite surface, the two dipoles add to create a field over the film. This surface contribution to the flexoelectric effect is very sensitive to the surface structure and can have a different sign when the surface termination changes by just a single atomic layer[4]. We could consider the situation where this is the dominant mechanism, i.e., that the surface piezoelectric effect generates a constant electric field across the film that displaces the atoms.

Consider such a field **E**. To first order, this will result in a force on each sublattice κ (to first order) of $F_{\kappa\alpha} = e\, Z^*_{\kappa,\alpha\beta}\, E_\beta$. Then, we can write the resulting displacements as: $\delta r_{\kappa\alpha} = \widetilde{\Phi}_{\kappa\alpha,\kappa'\lambda} F_{\kappa'\lambda}$, where $\widetilde{\Phi}$ is the pseudo-inverse of the dynamical matrix. Consider the field is normal to the film: $E_z$. If the Born effective charge tensor is diagonal, then the force is in $\hat{z}$, i.e., $F_{\kappa z} = e\, Z^*_{\kappa,zz}\, E_z$ and $\delta r_{\kappa\alpha} = e\widetilde{\Phi}_{\kappa\alpha,\kappa'z} Z^*_{\kappa',zz} E_z$. For SrTiO$_3$, for a field of 0.001 Ha/Bohr, we find (normalized to Sr so $\delta r_{Sr,z} = 0$): $\delta r_{Ti,z} = 1.21$ Å, $\delta r_{O1,z} = -0.83$ Å, $\delta r_{O2,z} = -0.83$ Å, $\delta r_{O3,z} = -0.55$ Å. As the calculated values do not show the same trend in the displacements of the O and Ti sublattices observed experimentally, surface piezoelectricity is likely not a dominant factor in our case.

### iii) Electrostatic boundary conditions

Finally we consider the effect of electrostatic boundary conditions. The above discussion was assuming zero electric field boundary conditions, i.e., "closed circuit." In other words, we assume that, e.g., surface adsorbates screen any macroscopic electric fields through the material. If this is not the case, and we truly have "open circuit" boundary conditions (i.e., zero displacement field over the film), the picture may change significantly[4,5]. We can convert the results calculated with closed circuit boundary conditions using the methodology discussed in

Springolo et al[5]. The overall flexoelectric coefficients are simply related by a factor of the dielectric function, though it is somewhat more complicated to determine the specific components and the resulting displacements. Doing the exercise, we obtain the results in Table II. Overall, the displacements are significantly suppressed and do not match the experimental observations. This is because the polar phonons are significantly hardened due to depolarizing field present under open circuit boundary conditions. We also lose the qualitative difference in magnitude between Ti and O displacements. Thus we conclude that considering open-circuit boundary conditions cannot explain the experimental observations.

|  | Ti | O1 | O2 | O3 |
|---|---|---|---|---|
| $\varepsilon_{xx,z}$ | -0.002 | 0.002 | 0.001 | 0.005 |
| $\varepsilon_{zz,z}$ | -0.015 | 0.013 | 0.009 | 0.009 |
| $\varepsilon_{eff,z}$ | -0.002 | 0.002 | 0.003 | 0.002 |

**TABLE II**. Atomic displacements in Å in the z direction with respect to the Sr atoms in SrTiO$_3$ under open-circuit boundary conditions for transverse, longitudinal, and effective bending strain gradients with magnitude 0.001 Å$^{-1}$.

**Supplementary Text 4 - Ti 3$d$ band splitting in response to strain gradient and polarization**

In cubic perovskites, the degeneracy of the 3$d$ orbitals of the transition metal is broken due to the octahedral geometry of the surrounding O atoms, splitting into t$_{2g}$ and e$_g$ orbitals. In the bent SrTiO$_3$ case, the t$_{2g}$ and e$_g$ orbital degeneracy is further broken by the presence of the strain gradient and flexoelectric polarization. Using DFT calculations, Susarla et al.[6], studied how a tetragonal distortion and a polar distortion individually affects the t$_{2g}$ and e$_g$ band splitting in a lead titanate perovskite system. Following the same recipe, we discuss the ordering of the 3$d$ bands in our sample, with the extra subtlety that the axes of tetragonal distortion and polarization are now orthogonal and not parallel to each other.

In Supplementary Fig. 12 below, we show a schematic showing how the $t_{2g}$ and $e_g$ bands split in response to polarization in the out-of-plane z-direction or strain in the in-plane x-direction. The change in energy levels can be understood by seeing how each distortion affects the different Ti-O bond lengths. For example, compressive strain along the $x$-axis results in a decrease in the in-plane Ti-O bond lengths, resulting in an increase in the energy of the xy and xz orbitals with respect to the yz orbital for $t_{2g}$. Similarly, the energy of the $3x^2$-$r^2$ orbital is raised with respect to the $y^2$-$z^2$ orbital for $e_g$. We note that the sublevels from $e_g$ level splitting for the polarization and strain are described in different bases as the axis of distortion is distinct for the two cases.

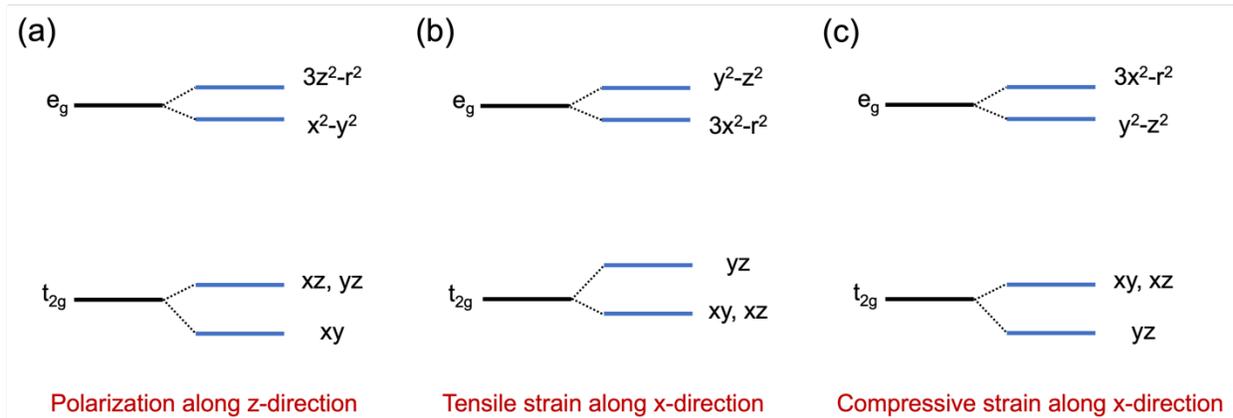

**Supplementary Fig. 12** | Schematic illustrating qualitatively the splitting of the Ti 3$d$ bands in response to (a) polarization along the z-axis, (b) tensile strain along the x-direction, and (c) compressive strain along the x-direction.

Combining the trends above, a qualitative scheme for the splitting of the Ti 3d bands at the top and bottom of the wrinkle is shown in Supplementary Fig. 13. Here, we assume that the strain has a much larger impact on the band splitting in comparison to the polar distortion. The band splitting shown here and the energy differences in the schematic are purely qualitative. The Ti $L_{2,3}$ EELS edges are sensitive to these changes in the Ti 3d energy levels – we do not resolve the individual peaks due to the ~1 eV energy resolution of our probe. Even with adequate energy resolution, the lifetime broadening of the peaks might prevent resolution of individual peaks.

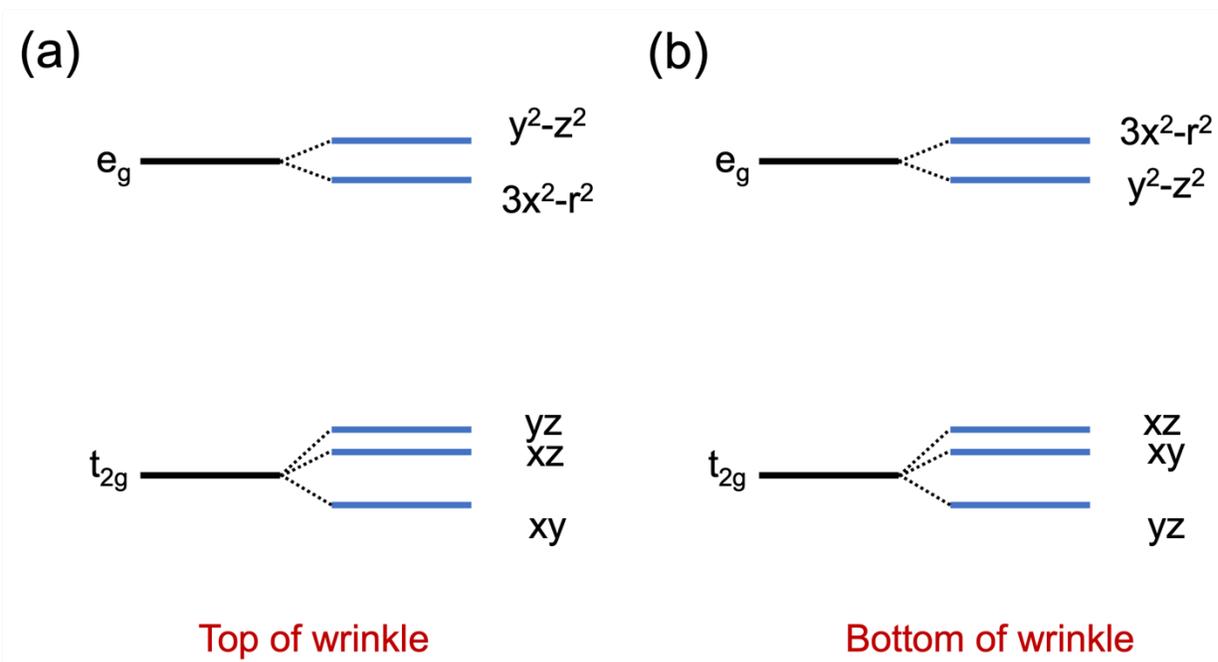

**Supplementary Fig. 13** | Schematic illustrating qualitatively the splitting of the Ti 3d bands expected at (a) top of the wrinkle (tensile strain in x-direction and polarization in z-direction), and (b) bottom of wrinkle (compressive strain in x-direction and polarization in z-direction).